\documentclass{icrc29}
\usepackage{graphicx,amssymb,amsmath,times}
\begin{document}
\title [Multifractal nature ...]{Multifractal nature of extensive air showers}
\author[A.Razdan]{ A. Razdan \\
                   Nuclear Research Laboratory,Bhaba Atomic Research Centre,\\
                   Mumbai 400 005, India }
        
\presenter{Presenter: A. Razdan (akrazdan@apsara.barc.ernet.in), \
ind-razdan-A-abs3-he15-poster}

\maketitle

\begin{abstract}
Cherenkov images are  multi-fractal in nature.
We show that multi-fractal behaviour of Cherenkov images 
arises  due to multiplicative
nature of pair production and bremsstrahlung processes
in the longitudinal shower development passage.
\end{abstract}

\section{\bf Introduction:}
                                       
Extensive Air Showers (EAS) [1] are produced when VHE/UHE primary
photons ,protons and other high Z nuclei enter atmosphere from the
top and produce Cherenkov radiation which can be studied by imaging
technique or by measuring lateral distribution. Cherenkov  images
recorded by a TACTIC-like $\gamma$-ray telescope represents the two
dimensional distribution of Cherenkov light pattern produced in the
atmnosphere. In the Hillas parameterization approach, image is
approximated as a ellipse and parameters like image "shape" and
"orientation" are calcualted. Hillas paramcters like Length, Width
Miss, Alpha, Distance ,Size etc are are essentially a
set of second order moments. Among these parameters
'$\alpha$ ' and 'Width' are known to be powerful
parameters for segregation
of $\gamma$-ACE from cosmic-ACE background. Using this approach,
present day telescopes have been able to reject cosmic ray background
events at 98 $\%$ level, while retaining up to 50$\%$ of the
$\gamma$-ray events from a point source. 
Hillas parameters are found to be good
classifiers for small images( close to telescope threshold energy)
but tend to fail for large images (of higher primary energy of the order
of 10 Tev or higher) as too many tail pixels are part of the image.
Again it is diffcult to study fainter compact sources and $\gamma$-rays
coming non-compact sources or of a diffusive origin. Up to 50$\%$
loss of actual image due to parameterization is also a big constraint.

It is interesting to observe that various products of EAS, like Cherenkov
photons and  particle density distributions ( on the ground) have shown
multifractal features. Multifractal nature of density fluctuations of 
particles has been experimentally verified and Lipshitz-Holder exponent
distribution of EAS has been found to be sensitive parameter to indentify
the nature of individual EAS [2]. We have shown that high energy $\gamma$-ray
simulated images of TACTIC [3] like $\gamma$-ray telescope are multi-fractal
in nature [4,5]. Multifracal nature of Cherenkov images have been confirmed
experimentally [6,7]. Because of multifractal nature of EAS products,
it becomes important to understand the  basic multifractal process 
that takes place during the longitudinal passage of EAS.

Fractals are self-similar objects which look same on many different
scales of observations.
Fractals are defined in terms of Hausdroff-Bescovitch
dimensions. Frcatal dimensions characterize the
geometric support of a structure
but can not provide any information about a possible distribution
or a probability that may be part of a given structure. This problem
has been solved by defining an inifinite set of dimensions known as
generalised dimensions which are achieved by dividing the object under
study into pieces ,each piece is labelled by an index i=0,1,2....N. If
we associate a probability $p_i$ with each piece of size $l_i$ than
partition function can be defined as
\begin{equation}
\Gamma(q,\tau)=\lim_{l \rightarrow 0} \Gamma(q,\tau,l)
\end{equation}
where
\begin{equation}
\Gamma(q,\tau,l)=\sum_{i=1}^{n} \frac{ p_i^{q}}{l_i^{\tau}}
\end{equation}
For unique function $\tau(q)$ it has been shown that
\begin{equation}
\Gamma(q,\tau)= \infty  
\end{equation}
for$ \tau < \tau(q) $
\begin{equation}
\Gamma(q,\tau)= 0 
\end{equation}
for $\tau > \tau(q)$.

This permits to define generalised dimension $D_q$
\begin{equation}
(q-1)D_q  =\tau(q)
\end{equation}
Here q is a parameter which can take all values between -$\infty$ to $\infty$.
This formalism is called as multi-fractal formalism which characterizes
both the geometry of a given structure and the probability measure
associated with it.

Cherenokov images 
arise due to EAS and have been found to be multifractal in nature.
In this paper we will attempt to understand underlying process 
responsible for multifractal behaviour of Cherenkov images.

\section{ Shower development as a multifractal process}

A ultra relativistic $\gamma$-ray enters atmosphere from the top and
interacts with air molecule to produce electron-photon cascade .  The
radiation length(x) for pair production and bremsstrahlung process is equal
in UHE/VHE region [1,10]. Particle -photon cascade in the atmosphere is
sustained alternately by electron ($e^{-}$) -positron ($e^{+}$) pair
production and bremsstrahlung process till the average energy per
particle  reaches critical value $E_c $ below which energy loss process
is mainly dominated by ionization process.

Shower development  can be visualized  a process in which a $\gamma$-ray
of energy $E_0$ (= 1 Tev (say)) after traveling distance x (on average)
produces electron-positron pair each having energy $\frac{E_0}{2}$. Since
energy is getting divided into two equal parts , we can attribute a resolution
of energy E =$2^{-1}$. In the next radiation length  both electron and
positron lose  half of their energy  (on average) and each radiates one
photon. Thus in this radiation length there are two photons and two particles
($e^{-1}$ and $e^{+1}$) each having energy $\frac{E_0}{4}$  which can
be attributed to energy resolution  E= $2^{-2}$. At this stage fraction
of photons ( $p_1$ =$\frac{1}{2}$ ) is same as fraction of particles
( $p_2$= $\frac{1}{2}$). As the shower develops into next radiation length
both electron and positron lose  half of their energy  and produce one
photon each. At the same time two photons produced in the previous radiation
length interact with air molecule to produce electron ($e^{-}$) -positron
($e^{+}$) pair. Thus in third radiation length  there are two photons
and six particles each having energy $\frac{E_0}{8}$ . This stage can be
attributed to the energy resolution E=$2^{-3}$. It is important to note
that, at this stage the process of equal division in energy between each
photon and each particle continues but the process of unequal measure between
photons and particles begins. In this radiation length the fraction of
photons is $p_1$=$\frac{1}{3}$ and fraction of particles (electron +positrons)
is $p_2$=$\frac{2}{3}$. The fourth radiation length corresponds to energy
resolution E=$2^{-4}$ as total of 16 photons and particles are produced
each having energy $\frac{E_{0}}{16}$. However, there are 10 charged particles
(5 positrons + 5 electrons ) and 6 photons, a case of unequal energy division and
unequal fraction.

At a distance of nx , the total number of particles and photons is
$2^{n}$ , each having average energy $\frac{E_{0}}{2^{n}}$ and on an
average shower consists of fraction of $\frac{2}{3}$ particles and
$\frac{1}{3}$ photons even at nth stage. This corresponds to energy
resolution of E=$2^{-n}$. At nth stage each particle or photon  can be
labelled sequentially  with i=0,1,2,..... The probability or fraction
of particles and photons can be written as $p_{i}$=$p_1^{k} p_2^{n-k}$.
The partition function for finite energy can be written as
\begin{equation}
\Gamma(q,\tau,E)=\sum_{i=1}^{n} \frac{ N_k p_i^{q}}{E_i^{\tau}}
\end{equation}
where $N_k$ is the number of particles and photons.
For the simplicity of calculations, we assume that incoming energy E is
equal to unit energy. This will not make any difference to actual results
but integrate it with other classical examples of multifractal behaviour,
e.g. curdling of cantor set [9]. On the average shower development process
has a recursive structure similar to cantor set [9] because both processes
are inherently binomial multiplicative in nature. For $\tau$=$\tau(q)$,we have
\begin{equation}
\Gamma(q,\tau(q),E)=1.
\end{equation}
In the limit of E $\rightarrow$ 0, the most dominant contribution to this
partition function will survive when $\tau$=$\tau(q)$, where $\tau(q)$ is the
solution of the equation
\begin{equation}
(p_1 ^{q} E ^{-\tau(q)}+p_2 ^{q} E^{-\tau(q)})^n =1
\end{equation}
Above equation can be easily solved to get $\tau(q)$.

\section{Case of equal energy and unequal fractions}

In each radiation length
the measure of photons and particles fluctuates  but on the average the shower
consists of $\frac{2}{3}$ positrons and electrons and $\frac{1}{3}$ photons.
In the present case E= $\frac{1}{2}$, $p_1$= $\frac{1}{3}$, $p_2$=$\frac{2}{3}$.
Using above equation and Stirling approximation, we have
\begin{equation}
D_q= \frac{1}{q-1}  \frac {ln(p_1^q+p_2^q)}{ln E}
\end{equation}
For q=0, $D_0$ gives fractal dimension. For q=1, $D_1$ is the information
dimension which encodes the entropy scaling  and q=2, $D_2$ is the
correlation dimension which measures scaling of two point density correlation.
Apart from $D_0$,$D_1$ ,$D_2$ there are infinite set of other exponents
from which information can be obtained by constructing an equivalent
picture of the system in terms of scaling indices '$\alpha$' for the
probability measure defined on a support of fractal dimension f($\alpha$).
f($\alpha$(q)) is the fractal dimension of the set. In case all $D_q$'s
are equal to fractal dimensions $D_0$ than f($\alpha$(q)) collapses to
a point f= $\alpha$ = $D_0$ ,indicating that scaling behaviour of a
fractal measure is of single type. This is situation which involves neither
probability variation nor length variation.

\section { Case of unequal energy and unequal fractions}

It is possible that as the shower develops deep into atmosphere, division of
energy may not remain equal. With the possibility of $E_1 \neq E_2 \neq \frac{1}{2}$ and/or
$p_1 \neq \frac{1}{3}$ and $p_2 \neq \frac{2}{3}$ shower development process can still be 
described as a multifractal process because either the energy or the probability or both 
need to be different for a multifractal process. The partition function for this case
can be written as
\begin{equation}
(p_1 ^{q} E_1 ^{-\tau(q)}+p_2 ^{q} E_2^{-\tau(q)})^n =1.
\end{equation}
where $E_1 + E_2$=E and $p_1 +p_2$ =p. Above equation can be numerically solved for
$\tau(q)$ if values of $E_1$,$E_2$,$p_1$ and $p_2$ are known and fixed for each radiation
length.

\section {Case of loss of energy}

As the shower develops deep into atmosphere most realistic situation is that there will be
energy loss in all radiation lengths. So a generalized scenario can be described in which 
$E_1 \neq E_2 \neq \frac{1}{2}$ and 
$p_1 \neq \frac{1}{3}$ and $p_2 \neq \frac{2}{3}$ with the condition $E_1 +E_2 < $  E in each radiation 
length. Each shower is unique because shower development process is random  and there are no unique
values of $E_1$, $E_2$, $p_1$ and $p_2$. 

\begin{figure}
\begin{center}
\includegraphics*[width=0.8\textwidth,angle=270,clip]{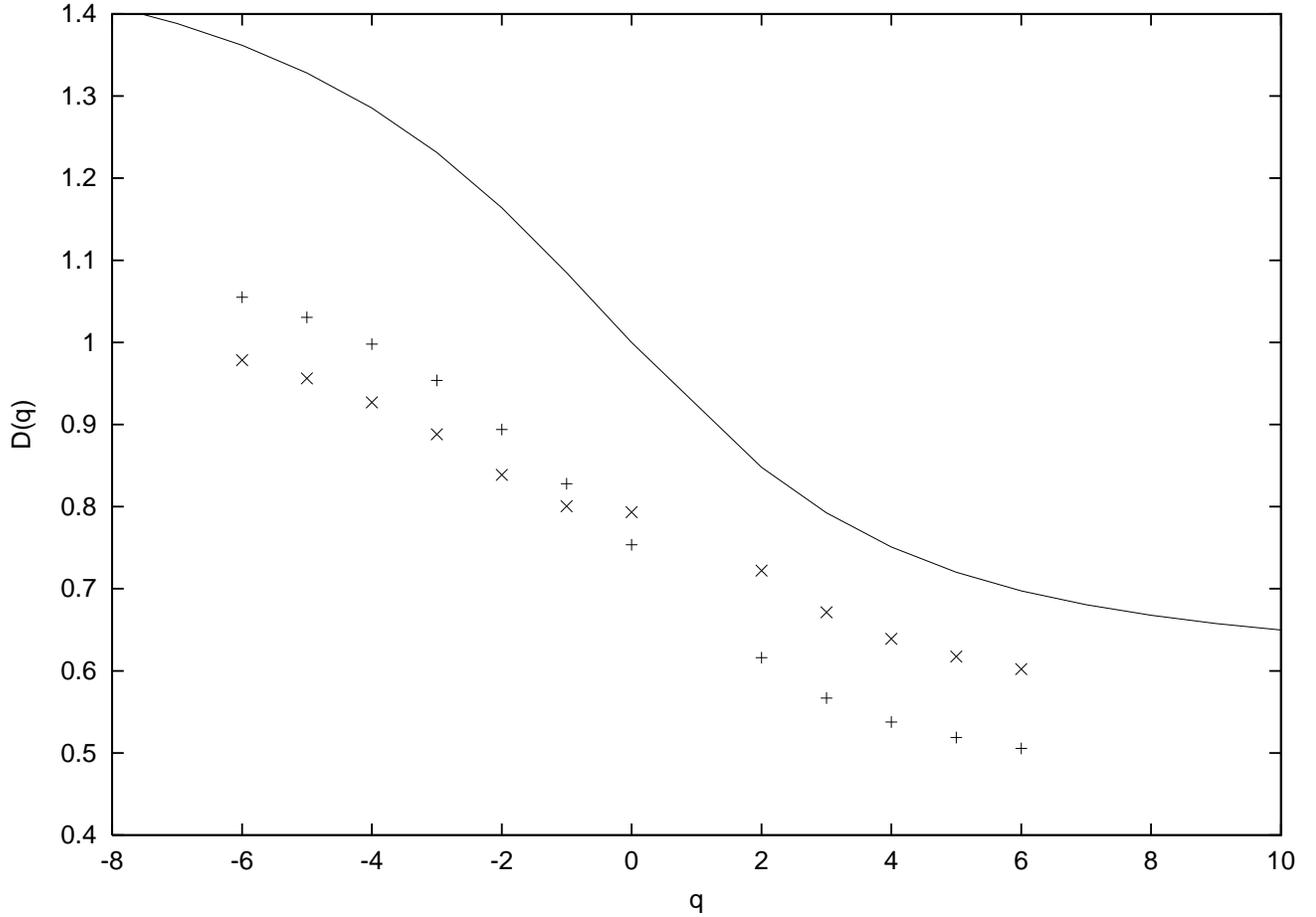}
\caption{\label {fig1} Variation in the muon (4-layer) rate over}
\end{center}
\end{figure}

\section {Simulation studies}

Any simulation / experimental arrangement to measure EAS will consist of set of detectors which
can measure only a sample or fraction of EAS products. This sample may consist of distribution of
charged particles / photons. The set of detectors which measure this distribution represents a 
geometric support and multifractal measures can be related to 
this geometric support. 
This can be done by calculating multifractal moments of this distribution
and obtain generalized dimensions from scaling properties of multifractal moments.
The definitions of multifractal properties given in previous section
are not defined with respect to any support and hence
can not be applied
directly to simulated / experimental data.

Simulated cherenkov images were generated for $\gamma$-rays and protons using CORSIKA code
for TACTIC configuration [4,5 and references therein]. $\gamma$-rays of energy 50 TeV and protons of energy 100 TeV are
considered for multifractal studies. Each simulated image is divided into M= $2^{\nu}$ where
$\nu$=2,4,6,8, is the scale. The multifractal moments are 
\begin{equation}
G_q= \sum_{i=1}^{M} (\frac{k_j}{N})^q
\end{equation}
where $k_j$ is the number of photoelectrons in the kth cell and N  is the total number of
photoelectrons in whole image. $G_q$ shows a power law behaviour with M, i.e.
$G_q= M^{\tau(q)}$ where
$\tau(q)$ is related to generalized multifractal dimension  by
\begin{equation}
D_q=\frac{\tau(q)}{q-1}
\end{equation}
We have calculated average values of $D_q$ for 1000 images each for
$\gamma$-rays and protons. We have repeated above studies for 30 TeV $\gamma$-rays and 60 TeV protons. No
energy dependence of $D_q$ on q was observed.

\begin{figure}
\begin{center}
\includegraphics*[width=0.8\textwidth,angle=270,clip]{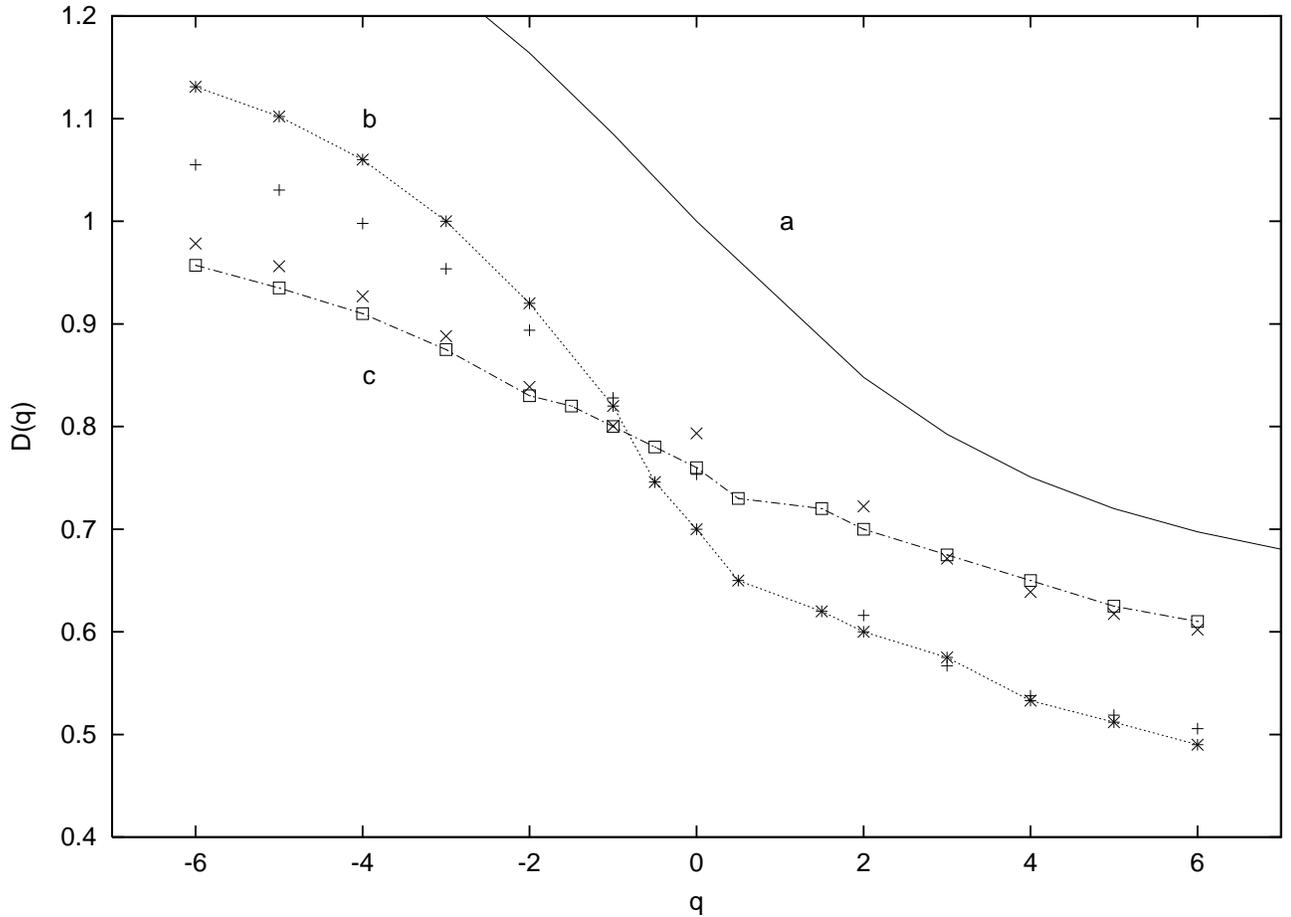}
\caption{\label {fig1} Variation in the muon (4-layer) rate over}
\end{center}
\end{figure}

\section{Results:}

As the shower develops deep into the atmosphere, values of $E_1$,$E_2$,$p_1$ and $p_2$ in various radiation lengths
may not remain fixed but may fluctuate. It is not possible to solve equation (4) for all fluctuating values in
each radiation length. Hence we have to take average values. Figure 1 depicts $D_q$ dependence on q for various 
average values. Continuous curve has been obtained in figure 1  for the case $E_1$=$E_2$=$\frac{1}{2}$,
$p_1$=$\frac{1}{3}$ and $p_2$=$\frac{2}{3}$. Simulated data corresponding to protons is represented by (+) sign
and simulated data corresponding to $\gamma$-rays is represented by (x) sign. $D_q$ values of each cherenkov
image corresponding to $\gamma$-rays and protons were calculated for 1000 images and average was computed. In
figure 1, these average values of $D_q$ for both protons and $gamma$-rays have been shown. It is clear from
figure that theoretically known average values of shower progression does not match with average values of simulated
results. 

Figure 2 shows average values of $D_q$ versus q behaviour for five different cases. Curve 'a' is actually
part (of continuous curve) of figure 1, shown here for the purpose of comparison. Average values of $D_q$ corresponding
to proton (+ sign shown as curve d) and $\gamma$-rays ( x sign shown as curve e) have been replotted.
The continuous corresponding to 'b' and 'c' have been obtained . Curve 'b' corresponds to $E_1$=$\frac{13}{35}$,
$E_2$=$\frac{22}{50}$, $p_1$=$\frac{2}{5}$ and $p_2$=$\frac{3}{5}$. Curves 'b' correspond to 23 $\%$ loss of
energy in proton initiated showers in all radiation lengths on the average and curve 'c' corresponds to average loss
of 18 $\%$  energy in all radiation lengths by $\gamma$-ray initiated shower. Curve 'a' corresponds to case of no 
energy loss.
It is clear that theoretical model of shower development
considered with average loss of energy in each radiation length is a good approximation of showers.

\section{Discussion:}

In this paper we have theoritically  calculated generalized dimension $D_q$ of shower development process based
on simple model for the case of average values of energy energy and probability. We have considered the case
of no loss of energy and average loss of energy per radiation length in the shower development process. We have
also simulated Cherenkov images corresponding to $\gamma$-rays and proton initiated showers. Average values of
$D_q$ have been calculated for both types of images. Comparison of average theoritical values and average simulated
values are given in figures 1 and 2. It is clear that theoritical model of shower development considered with
average loss  of energy in each radiation length in this paper is good approximation for showers.

It is interesting to observe that initiated EAS progression comes close to the mathematical definition of classic
cantor set. Curves b and c in figure 2, represent two scale fractal measure of cantor set in which a fraction
of energy at each stage gets removed. Such processes have been discussed [15] in detail [page 86]. The identification
of EAS model as a multifractal process is actually experiencing a cantor process or cantor dust in actual
physical world. 

In EAS process there is division of energies in terms of charged particles and photons as the appearences of charged
particles, in any radiation length, means production of cherenkov radiation and appearence of photons in any
radiation length means absence of cherenkov radiation. The cherenkov image formed on telescope is actually 
superimposition of cherekov photons produced at different heights. Multifractal measures are associated the distribution
of data on a geometric support. This support may be ordinary plane surface, photomultiplier tube (PMT) camera
or a fractal itself.

For particle detection, it has been shown by Kempa[13] that multifractal structures of density fluctuations of charged 
particles near EAS can be used to distinguish between $\gamma$-ray and hadron content of cosmic ray.
The method of multifractal moment analysis has been applied to simulated data as observed by KASKADE experiment
to distinguish EAS lateral distributions due to various types of primaries [12]. The fractal moments are
part of the sample of observables used for KASKADE data anlysis [13].

Another important point that is clear from this studies  is  that so called a 'toy model' of EAS can match detailed
simulated CORSIKA results just by introducing a small loss term in each radiation lenght.

The different values of $E_1$, $E_2$, $p_1$ and $p_2$ fitted in figure 2 are not unique values. Many other
combinations of energies and probabilites will produce same resutls. It is because EAS process is random in
nature and all combinations of different values of energies and probabilites are equally and uniquely probable.
It is important to note that there cannot be any unique values ascribed to energies and probabilites in EAS.

\end{document}